\documentclass[a4paper,12pt,english]{article}
\usepackage{amsfonts}
\usepackage{graphicx}
\usepackage{amsmath}
\usepackage{color}

\newcommand{\be}{\begin{equation}}
\newcommand{\ee}{\end{equation}}
\begin{document}


\begin{titlepage}
\begin{center}

\noindent{{\LARGE{Time-dependent backgrounds from marginal deformations of Minimal Strings in AdS$_3$}}}

\smallskip
\smallskip

\smallskip
\smallskip
\smallskip
\smallskip
\smallskip
\smallskip
\noindent{\large{Eoin Dowd, Gaston Giribet}}

\smallskip
\smallskip

\smallskip
\smallskip

\smallskip
\smallskip
\centerline{Department of Physics, New York University}
\centerline{{\it 726 Broadway, New York, NY10003, USA.}}

\end{center}

\bigskip

\bigskip

\bigskip

\bigskip

\begin{abstract}
We study a class of time-dependent backgrounds in string theory which consist of marginal deformations of minimal strings on AdS$_3$. For such backgrounds, we compute the three-point amplitudes and analyze their properties.
\end{abstract}
\end{titlepage}


\section{Introduction}

In this paper, we will be concerned with the computation of observables in string theory on time-dependent backgrounds. We will study a class of time-dependent backgrounds constructed as marginal deformations of the string $\sigma $-model on Euclidean AdS$_3$ space in direct product with a $c<1$ Generalized Minimal Model and a compact space. The $c<1$ theory is to be described by a timelike Liouville direction, while the AdS$_3$ factor corresponds to the $H_3^+=SL(2,\mathbb{C})/SU(2)$ Wess-Zumino-Witten (WZW) model at finite Kac-Moody (KM) level $\kappa $. We will focus on 26-dimensional backgrounds of bosonic string theory with an AdS$_3\times \mathbb{R}$ factor, $\mathbb{R}$ being time. For such backgrounds, we will explicitly compute three-point string amplitudes and discuss their properties. Amazingly, for certain parameter values, these amplitudes take a remarkably simple form. This is due to cancellations that occur between the AdS$_3$ and the timelike Liouville contributions. These cancellations are reminiscent of those that take place in Minimal Liouville Gravity \cite{Zreloaded, 05, 052, KP, KP2} and in type IIB superstrings on AdS$_3\times S^3\times M_4$ \cite{Pakman, Gaberdiel, Nicolas}. As we will discuss, such simplifications are no coincidence and can be explained in terms of the Liouville-$H_3^+$ WZW correspondence \cite{RibaultTeschner, HikidaSchomerus}. We will consider a large class of marginal deformations of the AdS$_3\times \mathbb{R}$ background, some of which having the interpretation as string theory on non-isotropic cosmological models with constant-time slices of non-zero curvature. Other solvable deformations of this class can be thought of as time-dependent current-current deformations similar to those studied in connection to the single trace $T\bar{T}$-deformations of AdS$_3$/CFT$_2$. For marginally deformed backgrounds we will also discuss the computation of three-point amplitudes. While, in contrast to the undeformed case, the three-point amplitude in the presence of a marginal deformation does not necessarily take a simple form in terms of special functions, the Coulomb gas approach still suffices to produce a relatively succinct integral expression. We explain how to deal with the analytic continuation of it.

The paper is organized as follows: In section 2, we study the worldsheet conformal field theory; that is to say, the $\sigma$-model on the undeformed AdS$_3\times \mathbb{R}\times M_{D-4}$ background, its symmetries and its spectrum. In section 3, we discuss the three-point amplitudes on such spacetime. In section 4, we consider a large class of marginal deformations, discuss different particular cases, comment on their geometrical interpretation and compute the free field representation of their three-point amplitudes. Section 5 contains final remarks and, in particular, a discussion linking this work to other time-dependent backgrounds considered in the literature.

\section{Worldsheet CFT}

We will study backgrounds of the form
\begin{equation}
\text{AdS}_3\, \times \, \text{Liouville}_{c<1}\, \times \,  M_{D-4}, \label{Action}
\end{equation}
where $M_{D-4}$ represents a CFT on a ($D-4$)-dimensional compact manifold. The Polyakov action on Euclidean AdS$_3$ with NS-NS fluxes is equivalently described by the gauged $H_3^ +$ WZW action with KM level $\kappa $, provided that the AdS$_3$ radius is given by $R=(\kappa\alpha')^{\frac 12}$. In Wakimoto representation \cite{Wakimoto}, and after taking into account quantum corrections, the AdS$_3$ piece of the Polyakov action reads \cite{GKS, Ooguri, KS}
\begin{equation}
S_{\text{AdS}}=\int d^2z \,\Big( \,\frac 12 \partial \varphi \bar{\partial}\varphi -\sqrt{\frac{2}{k-2}}R\varphi +\beta \bar{\partial }\gamma +\bar\beta {\partial }\bar\gamma-\beta \bar\beta e^{-\sqrt{\frac{2}{k-2}}\varphi}\, \Big)\label{TRR}
\end{equation}
which involves a scalar field $\varphi $ with non-vanishing background charge, coupled to a dimension-$(1,0)$ $\beta$-$\gamma$ ghost system. By integrating out the auxiliary fields $\beta $ and $ \bar \beta$, one recovers the string $\sigma $-model on the Euclidean AdS$_3$ space with metric
\begin{equation}\label{metric}
ds^ 2 = R^ 2 \,(\,d\varphi^ 2 + e^{2\varphi }d\gamma d\bar\gamma \,)  
\end{equation}
and support from a $B$-field and a linear dilaton. The near boundary region of AdS$_3$ corresponds to large $\varphi $, where the theory becomes free. In terms of the free fields, the currents that generate the affine KM symmetry of the theory take the form
\begin{eqnarray}
J^+(z)&=&\beta(z)\label{current1}\\
J^3(z)&=&-\beta(z)\, \gamma(z)\,-\sqrt{\frac{\kappa -2}{2}}\,\partial \varphi (z)\label{current2}
\\
J^-(z)&=&\beta(z)\,\gamma^2(z)\,+\sqrt{2(\kappa -2)}\, \gamma(z)\,\partial\varphi (z)\, + \, \kappa \,\partial\gamma (z)\label{current3}
\end{eqnarray}
together with their anti-holomorphic counterparts. The correlators of these fields are
\begin{equation}
\langle \varphi (z)\, \varphi(0)\rangle =-\log |z|^2 \ , \ \ \ \langle \beta (z)\, \gamma (0)\rangle ={z}^{-1}\ , \ \ \ \langle \bar\beta (z)\, \bar\gamma (0)\rangle ={\bar z}^{-1}
\end{equation}
from which one can easily verify the $SL(2,\mathbb{R})_{\kappa }$ commutation relations by computing the operator product expansion (OPE)
\begin{eqnarray}
J^{+}(z)J^{-}(0)&\simeq &{{\kappa}}\,{z^ {-2}}-{2J^3(0)}\, {z}^ {-1}+...\\
J^{3}(z)J^{\pm }(0)&\simeq &\pm {J^{\pm }(0)}\,{z}^{-1}+...\\
J^{3}(z)J^{3}(0)&\simeq &-\frac{\kappa }{2}\,{z^{-2}}+...
\end{eqnarray}
where $``..."$ stands for regular terms.

The remaining factor in (\ref{Action}) corresponds to a timelike Liouville action with central charge $c<1$. This is often referred to as a Generalized Minimal Model, or more simply, as a $c<1$ Liouville field theory. Its action takes the form 
\begin{equation}
S_{\text{LFT}}=\int d^2z \,\Big( -\frac 12 \partial X^0 \bar{\partial}X^0 -{Q}RX^0 +4\pi \mu\, e^{\sqrt{2}qX^0}\, \Big)\label{tL}
\end{equation}
with $X^0$ being a Liouville field with the {\it wrong} sign kinetic term and a background charge $Q=q-q^{-1}$. The parameter $\mu $ is often called the Liouville cosmological constant and can be set to 1 by appropriately shifting the zero mode of $X^0$, namely
\begin{equation}
X^0\to X^0-\frac{1}{\sqrt{2}q}\log \mu \, ;
\end{equation}
doing so alters the effective string coupling as $X^0$ enters linearly in the dilatonic term. The timelike Liouville field $X^0$ will represent the time direction of our background. That is to say, the background charge and the self-interaction potential in (\ref{tL}) can be interpreted as dilaton and tachyon-like background fields evolving in time. The interpretation of timelike Liouville theories as time-dependent backgrounds has been studied in the literature; see \cite{Strominger1, Strominger2} and references therein and, in particular, see the recent \cite{Itzhaki1, Itzhaki2}. Over the last twenty years, the timelike Liouville theory and its observables have been studied in detail; see for instance \cite{Zreloaded, 05, 052, KP, KP2} together with \cite{Timelike}-\cite{TimelikeU} and references thereof.

For concreteness, we will consider $M_{D-4}$ to be a $(D-4)$-torus, namely $M_{D-4}=T^N$ with $D=N+4$. In this case, the central charge of the entire background turns out to be
\begin{equation}
c=1-6Q^2+\frac{3\kappa }{\kappa -2}+N\equiv26 \, ,
\end{equation}
where the first two terms in the middle correspond to the Liouville theory, the third to the AdS$_3$ part, and $N$ comes from the scalars on the torus.

A particularly special surface in configuration space, along which very interesting things occur, is
\begin{equation}
q^2=\frac{1}{\kappa -2}\, .\label{relationen}
\end{equation}
At this point, the observables of the theory take a remarkably simple form in terms of $\Gamma$-functions, a phenomenon that, as we will explain, can be understood in terms of the $H_3^+$ WZW-Liouville correspondence \cite{RibaultTeschner, HikidaSchomerus}. Hereafter, we will assume (\ref{relationen}).

Finally, we will consider the critical dimension $D=26$, which yields the values 
\begin{equation}
\kappa =4\, , \ \ \ \ \ \ \ \ q^2=\frac{1}{2}\, , \ \ \ \ \ \ \ \ N=22.\label{background}
\end{equation}
These conditions almost fix our background. The only additional prescription is that we will demand one of the circles of the $22$-torus to have a self-dual radius; see below. The choice (\ref{background}) means that the contribution of the AdS$_3$ space to the central charge is $6$ (i.e. it is highly curved) while the timelike Liouville theory turns out to have a central charge $-2$. Timelike Liouville with $c=-2$ was studied earlier in \cite{Nakayama}; here, we will make use of some of the peculiar features observed therein.

The formula for conformal dimension in the worldsheet, after the Virasoro constraints have been imposed, leads to the mass-shell condition for the string spectrum; namely
\begin{eqnarray}
h=-\frac{1}{2}{J(1+J)}+\frac{1}{\sqrt 2}\alpha (1+{\sqrt 2}\alpha)+\frac{\alpha'}{4}P_IP^I{+\text{N}}\equiv 1,
\end{eqnarray}
together with
\begin{eqnarray}
\bar h =-\frac{1}{ 2}{J(1+J)}+\frac{1}{\sqrt 2}\alpha ({1}+{\sqrt 2}\alpha)+\frac{\alpha'}{4}\bar{P}_I\bar{P}^I{+\bar{\text{N}}}\equiv 1.
\end{eqnarray}
${\text{N}},\bar{\text{N}}\in \mathbb{Z}_{\geq 0}$ are the left and right oscillator numbers. $I=1,2,...N$ labels the coordinates on $T^N$, namely
\begin{eqnarray}
P^I=\frac{n^I}{R_I}+\frac{\omega^IR_I}{\alpha'}\ , \ \ \
\bar{P}^I=\frac{n^I}{R_I}-\frac{\omega^IR_I}{\alpha'}
\end{eqnarray}
with $n^I,\omega^I\in \mathbb{Z}$; $R_I$ are the compactification radii of the torus, while $n^I$ and $\omega^I$ are the Kaluza-Klein momenta and the winding numbers, respectively. $J$ and $\alpha $ represent the momenta in the radial AdS$_3$ and in the Liouville directions, respectively; see (\ref{KLKLKL2})-(\ref{KLKLKL}) below. Normalizable states in AdS$_3$ and in the (analytically continued) Liouville theory are such that
\begin{equation}
J=-\frac 12 +i\sqrt{2}{P_r} \, , \ \ \ \alpha=-\frac{1}{2\sqrt{2}}+P_0,\label{normalizame}
\end{equation}
with $P_r\in \mathbb{R}_{\geq 0}$, $P_0\in i\mathbb{R}$. With this, the Virasoro condition yields 
\begin{equation}
    (P_r)^2 + (P_0)^2 + \frac{{\alpha'}}{4}P_I P^I = 1
\end{equation}

As mentioned, we will assume at least one of the $N$-torus cycles to be compactified with the self-dual radius $R_5=\sqrt{\alpha'}$. As it is well known, the current algebra for this cycle of the worldsheet theory gets enhanced to the holomorphic and antiholomorphic copies of
\begin{equation}
SL(2,\mathbb{R})_{\kappa }  \times U(1)^{N}\,  \to \, SL(2,\mathbb{R})_{\kappa } \times {SU}(2)_{\kappa '=1} \times U(1)^{N-1} ,
\end{equation}
with the $SU(2)_1$ affine symmetry being generated by the KM currents
\begin{eqnarray}
K^{3}(z)=\frac i2 \partial X^5{(z)} \ , \ \ \ K^{\pm}(z)=\exp \Big(\pm X^5(z)\Big),
\end{eqnarray}
along with the antiholomorphic counterparts (hereafter, we set $\alpha'=4$). This, along with the $U(1)^{N-1}$ currents $i\partial X^{I>5}(z)$ generate the full affine symmetry of the worldsheet CFT.

Compactification in the self-dual radius enables us to consider states with $P^{I> 5}=\bar P^{I> 5}=0$, and $n^5=\pm 2$, $\omega^5=0$ or $n^5=0$, $\omega^5=\pm 2$, yielding $P_IP^I=\bar{P}_I\bar{P}^I=1$. This will allow us to have the special relation $P_0=iP_r$ between the radial momentum in AdS$_3$ and the momentum along the timelike Liouville direction.

\section{Amplitudes}

For the theory introduced above, we will compute string amplitudes. These are given by correlation numbers on the product CFT. The tachyon-like vertex operators creating primary states with $\text{N}=\bar{\text{N}}=0$ in the worldsheet are of the form
\begin{equation}
\mathcal{T}_P (x,\bar x) \,=\, \mathcal{N}_P\, \int d^2z\, V_{J}(x,\bar x|z,\bar z)\times e^{\sqrt 2 \alpha X^0(z,\bar z) + i\sqrt{2}P_IX^I(z,\bar z)},\label{KLKLKL2}
\end{equation}
where
\begin{equation}
V_{J}(x,\bar x|z,\bar z) = \frac {1}{\pi}\, \Big(\, |x(z)-\gamma (z) |^2\,e^{\sqrt{\frac{2}{\kappa -2}}\varphi (z,\bar z)} + e^{-\sqrt{\frac{2}{\kappa -2}}\varphi (z,\bar z)}\Big)^{-2(J+1)}\label{KLKLKL}
\end{equation}
are normalizable wave functions on $H_3^+$, provided $P_r\in \mathbb{R}$ in (\ref{normalizame}). The complex variable $x$ is commonly used to organize the vectors of the $J$-representation of $SL(2, \mathbb{R})$ and is usually referred to as the $x$-basis, in contrast to the $M$-basis which we will use in section 5. The latter is related to the former by a Mellin transform. The contribution to these vertices from the timelike Liouville theory and $N$-torus takes the usual exponential form.

$S$-point scattering amplitudes in the background (\ref{Action}) are thus given by correlation numbers 
\begin{equation}
\mathcal{A}^{S}_{P_1,P_2,...P_S}\, =\, \int\prod _{i=4}^S d^2z_i\, \left<\prod _{i=1}^S \,:\mathcal{T}_{P_i} (x_i):\,\right>
\end{equation}
where we are omitting the $c\bar c$ ghosts contributions. The subindices $P_i$ in the vertices collectively refer to the momenta $J_i ,\alpha_i , P_i^I$ of the $i^{\text{th}}$ state. $\mathcal{N}_{P_i}$ in front of each vertex stands for a $P_i$-dependent normalization that can be conveniently chosen. Notice that we are considering the three first vertices ($i=1,2,3$) to be inserted at fixed points. This suffices to cancel the infinite volume of the conformal Killing group, $PSL(2,\mathbb{C})=SL(2,\mathbb{C})/Z_2$. Projective invariance in the worldsheet enables us to set $z_1=0$, $z_2=1$ and $z_3=\infty $. Analogously, we set $x_1=0$, $x_2=1$ and $x_3=\infty $.

The result for the three-point amplitude ($S=3$) can be written explicitly. It takes a factorized form
\begin{equation}
   \mathcal{A}^3_{P_1,P_2,P_3} = \mathcal{N}_{P_1}\mathcal{N}_{P_2}\mathcal{N}_{P_3}\, C^H_{\kappa}(j_1, j_2, j_3) \,C^L_Q(\alpha_1, \alpha_2, \alpha_3)\, \delta^{N}(P_1 + P_2 + P_3)\;,
\end{equation}
where $ C^H_{\kappa }(j_1, j_2, j_3)$ and $ C^L_Q(\alpha_1, \alpha_2, \alpha_3)$ are the structure constants of the $H_3^+$ WZW model \cite{Teschner1,Teschner2,Teschner_2001} and the timelike analog of the Dorn-Otto-Zamolodchikov-Zamolodchikov (DOZZ) structure constants of Liouville theory \cite{DO,ZZ}, respectively. The latter have been computed in \cite{Zreloaded, KP, KP2, Timelike, HMW, Giribet}. The $N$-dimensional delta function contribution comes of course from the torus. The explicit form of the structure constants above is known; for generic $\kappa $ and $q$, they have an abstruse form in terms of special $\Upsilon$-functions \cite{ZZ}, cf. \cite{Barnes, Ponsot:2003ju}. A remarkable simplification occurs for backgrounds with special values for $\kappa,\, q$ and $N$. Using the shift symmetry of the $\Upsilon$-functions and the specific values (\ref{background}), the final result for the three-point amplitude takes the form
\begin{equation}\label{resultado}
        \mathcal{A}_{P_1,P_2,P_3}^3 \, = \, \frac{\Gamma \Big(\frac 14 -\frac{1}{\sqrt 2}\sum_{k=1}^3P_0^k\Big)}{\Gamma \Big(\frac 34 +\frac{1}{\sqrt 2}\sum_{k=1}^3P_0^k\Big)} \, \prod_{i=1}^3\,  \frac{\Gamma \Big(\frac 14 -\frac{1}{\sqrt 2}\sum_{k=1}^3P_0^k+\sqrt 2 P_0^i\Big)}{\Gamma \Big(\frac 34 +\frac{1}{\sqrt 2}\sum_{k=1}^3P_0^k-\sqrt 2 P_0^i\Big)}\, .
\end{equation}
In writing this, we have chosen the normalization of the vertices to be 
\begin{equation}
\mathcal{N}_{P_i} \,= \, (8\mu )^{\sqrt 2 P_0^i}\,\frac{\Gamma (1+2\sqrt 2 P^i_0)}{\Gamma (-2\sqrt 2 P^i_0)}\, ,
\end{equation}
for $i=1,2,3$. In general, the Knizhnik-Polyakov-Zamolodchikov (KPZ) scaling of the three-point correlators would be 
\begin{equation}
\sim \, \mu ^{\frac 12 -\sqrt 2 \sum_{k=1}^3P_0^k}\, ;
\end{equation}
by shifting the zero mode of $X^0$, we can set $\mu=1/8$ for convenience. 

It may come as a surprise to anyone familiar with Liouville field theory that (\ref{resultado}) does not involve any $\Upsilon$-function. This is due to a notable cancellation between the Liouville and the WZW structure constants. Therefore, the pole structure of the three-point amplitudes is easily read from the dependence of $\Gamma$-functions in the numerator of (\ref{resultado}). The expression exhibits poles at
\begin{equation}
\frac{1}{\sqrt 2}\sum_{k=1}^{3}P_0^k \in \mathbb{Z}_{\geq 0} +\frac 14 \, , \ \ \ \ \ \frac{1}{\sqrt 2}\sum_{k=1}^{3}P_0^k- 
\frac{2}{\sqrt 2}P_0^i \in \mathbb{Z}_{\geq 0}  +\frac 14
\end{equation}
with $i=1,2,3$. It is also evident from (\ref{resultado}) that the amplitude is manifestly invariant under crossing 
\begin{equation}
P_0^{i}\, \to \, P_0^{j}\, ,\ \ \ \ i,j=1,2,3.
\end{equation}

As said in the introduction, the fact that the three-point function (\ref{resultado}) takes such a simple form in terms of $\Gamma$-functions with no need of $\Upsilon $-functions is reminiscent of the cancellations that take place in Minimal Liouville Gravity \cite{Zreloaded, KP, KP2}, in the 2D Virasoro string \cite{Eberhardt}, and in the IIB superstring amplitudes of $1/2 $ BPS states in AdS$_3\times S^3\times M_4$ \cite{Pakman, Gaberdiel}. As we will explain, this is not a coincidence but a manifestation of the $H_3^+$ WZW-Liuouville correspondence at the special point $\kappa =4$. Still, it is quite surprising that a model describing a time-dependent, curved string background at finite curvature $R^2/\alpha '$ can be solved explicitly and yields such a manageable analytic form. 

In the next section, we will consider marginal deformations of the background (\ref{Action})-(\ref{background}) that describe a more general class of time-dependent string backgrounds, and we will consider the three-point functions on those spaces.

\section{Marginal deformations}

Let us study marginal deformations of the background (\ref{Action}). First, we can consider operators that, apart from preserving conformal symmetry, are also invariant under the affine symmetry $SL(2,\mathbb{R})_{4}$. Operators
\begin{equation}
\delta S=\int d^2 z\,\beta(z) \bar \beta(\bar z)\, e^{-\varphi (z,\bar z)-X^0 (z,\bar z)}
\Big (\, \lambda_1 \, +\,\lambda_2\, \beta(z) \bar \beta(\bar z)\, e^{-\varphi (z,\bar z)}\Big)\label{deforme}
\end{equation}
are of that sort. Notice that the case $\lambda _2 =0$ is the interaction term in the WZW action dressed with a dimension-($0,0$) Liouville operator. Such an operator describes a gravitational coupling in the worldsheet action, as it can easily be made manifest by bosonizing the $\beta $-$\gamma$ system. Operators with $\lambda_2\neq 0$ have less clear interpretation in the string $\sigma $-model as, after bosonization, they seem to correspond to higher-derivative couplings. The latter operators are usually known as the {\it second screening} operator and have been studied in detail in \cite{GN3}. 

Of course, there are also operators that, while being marginal in the sense that they respect worldsheet conformal symmetry, do not necessarily commute with the full set of KM currents (\ref{current1})-(\ref{current3}). One such operator is
\begin{equation}
\delta \hat S= \int d^2 z\, e^{-\varphi (z,\bar z)}
\Big (\, \hat\lambda_1 \, e^{X^0 (z,\bar z)} +\,\hat\lambda_2\, e^{-2X^0 (z,\bar z)}\Big)
\end{equation}
which also consists of a dimension-($1,1$) operator in the WZW theory dressed with dimension-($0,0$) operators in the timelike Liouville theory, growing either in the past ($X^0\to -\infty$, $\hat\lambda_{2}\neq 0$) or in the future ($X^0\to \infty$, $\hat\lambda_{1}\neq 0$). However, for the remainder of this section, we will be concerned with a broader class of deformations.

A particularly interesting class of marginal operators are those that are linear in $\beta\Bar{\beta}$. As said, they have a clear geometrical interpretation in terms of a deformation of the target space. We can propose an operator of the form
\begin{equation}
    \delta \tilde{S} = -\int d^2 z\, \beta(z) \bar \beta(\bar z)\, e^{-\varphi (z,\bar z)} \Lambda[\varphi,X^0] 
\end{equation}
which turns out to be marginal provided
\begin{equation}\label{TheO}
    \Lambda[\varphi, X^0] = e^{\frac{1}{2}(\varphi - X^0)} \int dP\, \left( {\lambda}_{+}(P) e^{i\sqrt{2}P (\varphi + X^0)} + {\lambda}_{-}(P) e^{-i\sqrt{2}P (\varphi - X^0)}  \right)\,.
\end{equation}
This can be thought of as the Laplace transform of the couplings $\lambda_{\pm}(P)$. By integrating the auxiliary fields $\beta, \bar\beta $, and choosing the coupling constants such that $\Lambda [\varphi, X^0]> - 1$ for all values of $\varphi\in \mathbb{R}_{\geq 0 }$, $X^0\in \mathbb{R}$, we get
\begin{equation}
    \tilde{S} =\int d^2z \,\Big( \,\frac 12 \partial \varphi \bar{\partial}\varphi + \frac{e^{\varphi}\, \partial\bar \gamma \bar\partial \gamma }{1+\Lambda [\varphi,X^{0}]} +\frac 12 \eta _{AB}\, \partial X^A \bar{\partial } X^B +R\Phi [\varphi , X^0]
    +T[X^0]\,
     \Big) 
\end{equation}
where $A, B=0,1,2,...D-4$. This describes a string $\sigma$-model with a linear dilaton field
\begin{equation}
    \Phi [\varphi , X^0] = \frac{1}{\sqrt 2}X^0(z,\bar z)-\varphi (z,\bar z) \label{Dila}
\end{equation}
and an exponential time-dependent tachyon profile
\begin{equation}
T[X^0] = \frac{\pi}{2}\, \exp\Big({\frac{1}{\sqrt{2}}X^0(z,\bar z)}\Big)
\end{equation}
The metric function is also deformed by a time-dependent factor, namely
\begin{equation}
G_{\gamma\bar\gamma}[\varphi, X^0] = \frac 12\, {e^{\varphi}\,  }\Big(1+\Lambda [\varphi,X^{0}]\Big)^{-1}\, .\label{RRRRR}
\end{equation}

The growth of the Liouville barrier of the marginal operators present in the action must be chosen in such a way that it prevents the theory from exploring the strong coupling region due to the linear dependence in the dilaton (\ref{Dila}).

Operators (\ref{TheO}) with $\lambda_{\pm }\neq 0=\lambda_{\mp }$ and $\pm i\sqrt 2 P<-\frac 12 $ damp off at infinity faster than the AdS$_3$ screening in (\ref{TRR}). The divergence that (\ref{RRRRR}) exhibits at $\Lambda =-1$ is understood as the neutralization of that screening operator, for which the worldsheet CFT behaves as strongly coupled.

There are interesting special cases of the operator (\ref{TheO}) that we can discuss. For example, consider the particular case $\lambda_{\pm}(P)=\lambda\, \delta(P)$, which yields the marginal deformation
\begin{equation}
\delta S \,= \, 2\lambda \int d^ 2z\, \beta(z)\bar{\beta}(\bar z)\, e^{-\frac 12 \varphi(z,\bar z ) -\frac 12 X^0(z,\bar z )  }.\label{La339}
\end{equation}
In fact, this is a generalization of (\ref{deforme}), in the sense that it breaks a $SL(2,\mathbb{R})$ symmetry by dressing a gravitational operator in AdS$_3$ with a Liouville operator of non-zero conformal dimension.

A one-parameter deformation of the latter operator is obtained by considering in (\ref{TheO}) the coefficients $\lambda_{\pm}(P)=\lambda \, \delta\Big(P+\frac{i\omega }{\sqrt{2}}\Big)$. This yields the deformation
\begin{equation}
\delta S \,= \, 2\lambda \int d^ 2z\, \beta(z)\bar{\beta}(\bar z)\, e^{-\frac 12 \varphi(z,\bar z )+(\omega  -\frac 12) X^0(z,\bar z )}\,\cosh \Big(\omega  \, \varphi(z,\bar z )\Big)
\end{equation}
for $\omega \in {\mathbb{R}}$; $\omega =0$ corresponds to (\ref{La339}). A sort of reciprocal case would be given by $\lambda_{\pm}(P)=\lambda \, \delta\Big(P\mp \frac{i }{2\sqrt{2}}\Big)$, which yields
\begin{equation}
\delta S \,= \, 2\lambda \int d^ 2z\, \beta(z)\bar{\beta}(\bar z)\, e^{- \varphi(z,\bar z )}\,\cosh \Big(X^0(z,\bar z )\Big)
\end{equation}

An interesting time-independent case is $\lambda_{\pm }(P)=t_{\pm }\,\delta \Big(P+\frac{i}{2\sqrt 2 }\Big)$. In this case, the deformation takes the form
\begin{equation}
\delta S \,= \, t_+ \int d^ 2z\, \beta(z)\bar{\beta}(\bar z)\,+ t_- \int d^ 2z\, \beta(z)\bar{\beta}(\bar z)\, e^{- \varphi(z,\bar z )}.\label{La42}
\end{equation}
This operator corresponds to a linear combination of the $J^+\bar{J}^ {+}$-deformation and the WZW screening operator. The particular case $t_+=0$ corresponds to the string $\sigma$-model on AdS$_3$, for which the adding of $t_-$ can easily be absorbed by shifting the zero-mode of the Wakimoto field $\varphi $ --at least for one of the signs of $1-t_-$. The latter renormalizes the effective coupling in AdS$_3$. The $J^+\bar{J}^ {+}$ operator for $t_+\neq 0$ is the one considered in the literature in the context of the single trace $T\bar T$-deformation; see \cite{TT1, TT2, TT3} and references therein and thereof. In this case, the string $\sigma $-model describes the theory propagating on a conformally flat manifold of non-constant curvature, which, while behaving as AdS$_3$ deep inside the bulk, asymptotes to flat space at infinity. 

Among the deformations (\ref{TheO}) we also find time-dependent solutions which describe homogeneous non-isotropic cosmologies that have hyperbolic (Euclidean AdS$_3$) constant-time slices at $X^0=X^0_*$. A necessary condition for this to occur is
\begin{equation}
\frac{d}{d\varphi }\Lambda [\varphi , X^0_*]=0\, .\label{Bianchi_condition}
\end{equation}
This yields
\begin{equation}
    \Lambda [\varphi, X^0] = \lambda_- +\lambda_+ e^{-X^0} + e^{\frac 12 ({\varphi - X^0})} \int dP\,\lambda(P) \, \sin(\sqrt{2} P (X^0 - X^0_*)) \,e^{i\sqrt{2}P\varphi} \,.\label{OPOPOP}
\end{equation}
Particular solutions of this class are $\lambda_{+}=0$, $p=- \frac{i}{2\sqrt 2}$ and $\lambda_{-}=0$, $p=+ \frac{i}{2\sqrt 2}$. One of these cases is the screening operator of the WZW theory (as mentioned, adding such operator results in a renormalization of the AdS$_3$ coupling \cite{Notes}), while the other case can be thought of as a time-dependent string coupling on AdS$_3$; namely, for $\lambda_-=0$, $\lambda_+=\lambda_1 \, \delta \Big(P-\frac{i}{2\sqrt 2}\Big)$, we get the operator
\begin{equation}
\delta S \,= \, \lambda_1 \int d^ 2z\, \beta (z)\bar \beta (\bar z )\, e^{- \varphi (z,\bar z )-X^0(z,\bar z )}\label{workingexample}
\end{equation}
where the $X^0$-dependence is given by the dressing of a dimension-$(0,0)$ Liouville operator ($\lambda_2=0$ in (\ref{deforme})). A similar example is $\lambda_+=0$, $\lambda_-=\lambda \, \delta \Big(P-\frac{i}{2\sqrt 2}\Big)$, which yields the time-dependent operator
\begin{equation}
\delta S \,= \, \lambda \int d^ 2z\,  e^{- X^0(z,\bar z )}\beta (z)\bar \beta (\bar z );
\end{equation}
that breaks the $SL(2,\mathbb{R})$ down to $U(1)$. This can be thought of as a time-dependent coupling to the $J^+\bar J^+$ deformation, cf. (\ref{La42}).

Returning to (\ref{OPOPOP}), we observe that for the field equations of the auxiliary fields $\beta, \bar\beta $ to be invertible and avoid the divergence in (\ref{RRRRR}), we require
\begin{equation}
\Lambda[\varphi, X^0] > -1 \label{La41bis}
\end{equation}
for all $\varphi\in\mathbb{R}_{>0}$, $X^0\in\mathbb{R}$. We need to be careful in picking the exponential dependence in such a way that the above expression is bounded below. One option is to let $P = -\pi \sqrt{2}\omega + \frac{i}{2\sqrt{2}}$ with $\omega\in\mathbb{R}$. With this, we have
\begin{align}
        \Lambda [\varphi, X^0] =& \lambda_{-} - e^{-\frac{X^0_*}{2}}\int_{\mathbb{R}} d\omega\,\tilde{\lambda}(\omega) e^{- 2\pi i\omega(\varphi + X^0 - X^0_*)}  \notag \\
        &+ e^{-X^0}\left[\lambda_+  + e^{\frac{X^0_*}{2}} \int_{\mathbb{R}} d\omega\,\tilde{\lambda}(\omega) e^{- 2\pi i\omega(\varphi - X^0 + X^0_*)}\right] \,,
\end{align}
and for $\Lambda >-1$, we can see that
\begin{equation}
    \lambda_+  + e^{\frac{X^0_*}{2}} \int_{\mathbb{R}} d\omega\,\tilde{\lambda}(\omega) e^{- 2\pi i\omega(\varphi - X^0 + X^0_*)} \geq 0\;,
\end{equation}
\begin{equation}
    \lambda_{-} - e^{-\frac{X^0_*}{2}}\int_{\mathbb{R}} d\omega\,\tilde{\lambda}(\omega) e^{- 2\pi i\omega(\varphi + X^0 - X^0_*)} > -1\;,
\end{equation}
are sufficient for $\Lambda>-1$. We have the two constants $\lambda_\pm$ and the function $\tilde{\lambda}(\omega)$ to play with. Note that the above integrals are nothing other than Fourier transforms of $\tilde{\lambda}(\omega)$. We can consider any integrable function $\tilde{\lambda}(\omega)$ provided it is bounded below. As an example, let us consider a Gaussian distribution with $\tilde{\lambda}(\omega) = e^{-\pi \omega^2}$. The corresponding Fourier transform
\begin{equation}
\label{eqn: example 1}
    \int_{\mathbb{R}} d\omega\,\tilde{\lambda}(\omega) e^{- 2\pi i\omega(\varphi \pm (X^0 - X^0_*))} = e^{-\pi (\varphi \pm (X^0 - X^0_*))^2}
\end{equation}
is also Gaussian. Moreover, since this distribution is bounded above and below, the above criteria are easily satisfied. Explicitly, by letting $\lambda_+ =0$ and $\lambda_- = e^{-{X^0_*}/{2}}$, we verify that both (\ref{Bianchi_condition}) and (\ref{La41bis}) are satisfied.

Now, let us consider amplitudes in the backgrounds with one of these marginal deformations turned on. This will enable us to analyze how the inclusion of the marginal operators alters the form of the amplitudes. As a working example, let us consider the operator (\ref{workingexample}), which we previously discussed in (\ref{deforme}). This is a $SL(2,\mathbb{R})_{4}\times SL(2,\mathbb{R})_{4}$ preserving operator dressed with a dimension-($0,0$) Liouville operator, which can be thought of as introducing a time-dependence in the AdS$_3$ effective coupling. We can take (\ref{workingexample}) to show how to compute string amplitudes in the presence of such a time-dependent deformation. In order to do so, let us consider the vertex operators
\begin{equation}
\mathcal{T}_{P,M,\bar M} \,=\, \frac{\Gamma (1-J+M)}{\Gamma (J-\bar{M})}\, \int d^2z\, V_{J,M,\bar M}(z,\bar z)\times e^{\sqrt 2 \alpha X^0(z,\bar z) + i\sqrt{2}P_IX^I(z,\bar z)}\;,
\end{equation}
which are defined in terms of the Mellin transform of operators (\ref{KLKLKL}); or more precisely,
\begin{equation}
V_{J,M,\bar M} (z,\bar z) = \int d^2x \, V_{J}(x,\bar x|z,\bar z) \, x^{J-M} \bar{x}^{J-\bar M}\; .
\end{equation}
The three-point amplitudes in the deformed theory are
\begin{equation}
\mathcal{A}^3_{P_1,P_2,P_3}\, =\, \left< \,:\mathcal{T}_{P_1,M_1,M_1}(0):\,\,:\mathcal{T}_{P_2,M_2,M_2}(1):\,\,:\mathcal{T}_{P_3,J_3,J_3}(\infty ):\, \,\right>_{\lambda _1}
\end{equation}
where $J_i=-\frac 12 +i\sqrt 2 P_0^i$, $i=1,2,3$, and where the subindex $\lambda_1$ makes it explicit that the expectation value is defined in the deformed theory. For simplicity, we are considering the deformation (\ref{deforme}) with $\lambda_2=0$; the generalization to arbitrary $\lambda_2$ is straightforward but lengthy. 

Resorting to the Coulomb gas representation of the correlation functions we arrive at the following integral expression for the amplitudes
\begin{eqnarray}\nonumber
\mathcal{A}^3_{P_1,P_2,P_3} = \sum_{n=0}^\infty \frac{\mu^{s_n}\lambda_1^{n}\Gamma (-s_n)\Gamma(-t_n)}{\Gamma (n+1)}\int \prod_{r=1}^{s_n} d^2w_r\, \Big[ \,  \prod_{r=1}^{s_n}|w_r|^{2\sqrt 2P_0^1-3 }|1-w_r|^{2\sqrt 2P_0^2-3}
\end{eqnarray}
\begin{eqnarray}\nonumber
\times \, 
\prod_{r'>r}^{s_n}|w_r-w_{r'}|^{-2}\int \prod_{j=1}^{n} d^2x_j\, \Big[ \,
\prod_{j=1}^{n} |x_j|^{2\sqrt 2 P_0^1-3}|1-x_j|^{2\sqrt 2 P_0^2-3}\int \prod_{l=1}^{t_n} d^2v_l\, \Big[ \,\prod_{l=1}^{t_n} |v_l|^{2\sqrt 2 P_0^1-1} 
\end{eqnarray}
\begin{eqnarray}\label{ultima}
\times \, |1-v_l|^{2\sqrt 2 P_0^2-1}
\prod_{l'>l}^{t_n} |v_l-v_{l'}|^{2} \prod_{r=1}^{s_n}\prod_{j=1}^{n} |x_j-w_r|^{-2}
\prod_{j=1}^{n}\prod_{l=1}^{t_n} |x_j-v_l|^{-2} \, \Big] \,\Big] \,\Big] \,
\end{eqnarray}
where 
\begin{eqnarray}
s_n= - t_n = -\frac 12 -n+\sqrt 2 \sum_{k=1}^3P_0^k \, .
\end{eqnarray}

The absence in (\ref{ultima}) of factors of the form $|x_j-x_{j'}|^{2\eta_{jj'}}$ and $|w_r-v_{l}|^{2\sigma_{jj'}}$ (i.e. the fact that $\eta_{jj'}=\sigma_{jj'}=0$) is a remaining feature of the neutralization between the Liouville and the WZW contributions to these observables. The effect of the marginal deformation is given by the factors $|x_j-w_r|^{-2}$ and $|x_j-v_l|^{-2}$.

Of course, the integral representation above is only defined for $s_n \in \mathbb{Z}_{\geq 0}$. In principle, this would limit the kinematic configurations for which (\ref{ultima}) is valid. However, the expression for generic $P_0^i$ can be obtained by analytic continuation $s_n \in \mathbb{Z}_{\geq 0}\to \mathbb{C}$. When trying so, one immediately faces another problem as $t_n=-s_n\in \mathbb{Z}_{\leq 0}$. To deal with this, we may resort to the formulae
\begin{equation}
\prod_{k=1}^{t_n}f(k) = \frac{1}{f(0)}\prod_{k=1}^{s_n -1}\frac{1}{f(-k)} \, , \ \ \ \ \ \prod_{k=1}^{s_n}f(k)\prod_{j=1}^{t_n}f(j) = \frac{f(s_n)}{f(0)}\prod_{k=1}^{s_n -1}\frac{f(\, k)}{f(-k)}
\end{equation}
which have been successfully used in similar CFT calculations; see \cite{Nicolas, Giribet, Dotsenko1, Dotsenko2, Dotsenko3, Rado}. Expression (\ref{ultima}) can be further simplified, e.g. by integrating in $v_l$ by resorting to the Fateev duality formula; see (18)-(21) in \cite{Fateev-B} and (1.13)-(1.14) in \cite{Fateev-L}. The resulting expression, however, is still cumbersome enough and not very illuminating; indeed, an expression in terms of simple functions for arbitrary $\lambda_1$ is not at hand. In contrast to the undeformed case ($\lambda_1=0$), the integrals can be explicitly solved in terms of $\Gamma$-functions. This yields a product of two Dotsenko-Fateev integrals of the form
\begin{eqnarray}
\int \prod_{r=1}^{s_0}d^2w_r\, \Big[\prod_{r=1}^{s_0}|w_r|^{2L_1}|w_r-1|^{2L_2}\prod_{r<r'}^{s_0}|w_r-w_{r'}|^{4\sigma} \, \Big]\, =\, \Gamma(s_0+1) \Big(\pi \frac{\Gamma (1-\sigma)}{\Gamma (\sigma)}\Big)^{s_0}\times \nonumber\\
\ \ \ \ \prod_{r=1}^{s_0} \frac{\Gamma (r\sigma) \Gamma (1+L_1+(r-1)\sigma ) \Gamma (1+L_2+(r-1)\sigma ) \Gamma (-1-L_1-L_2+(2-s_0-r)\sigma ) }
{\Gamma (1-r\sigma) \Gamma (-L_1+(1-r)\sigma ) \Gamma (-L_2+(1-r)\sigma ) \Gamma (2+L_1+L_2+(s_0+r-2)\sigma)};
\end{eqnarray}  
see (B.9) in \cite{DF}, cf. \cite{GL, BB}. This integral is defined over the entire $\mathbb{C}$-plane avoiding the divergences at the coincident points by means of an adequate prescription. This can be associated to multiple integrals of the Selberg type, and it can be thought of as a generalization of the Shapiro-Virasoro integral. After analytic extension of it and manipulation of $\Gamma$-functions one verifies that the result (\ref{resultado}) is recovered.

As a last example of a marginal deformation that admits an interpretation as a time-dependent string background, we can consider the sinh-Liouville non-gravitational operator
\begin{equation}
\delta S \,= \, 2\lambda \int d^ 2z\,  e^{- \varphi(z,\bar z )-\frac 12 X^0(z,\bar z )}\,\cosh \Big(\,\frac 32 X^0(z,\bar z )\Big)\, . \label{sinsh}
\end{equation}
The Coulomb gas techniques discussed above have also been applied in this case; see \cite{FukudaHosomichi, Giribet2022}. Other operators, like the $J^+\bar{J}^+$ deformation, have also been studied with similar techniques, cf. \cite{Giribet2018}.

\section{Discussion}

In this paper, we have studied time-dependent backgrounds of string theory of the form AdS$_3\times \text{Liouville}_{c<1}\times T^N$ and their marginal deformations. The time direction was described by a timelike Liouville field theory with negative central charge, while highly-curved Euclidean AdS$_3$ space appears as constant-time slices of 4-dimensional non-compact spacetime. For such background, we discussed the spectrum, the computation of three-point amplitudes, and a class of marginal deformations that admit a geometrical interpretation within the string $\sigma$-model. In the case of the undeformed background, and for certain parameter values, we have shown that the three-point amplitudes take a remarkably simple form. This is due to cancellations that occur between the AdS$_3$ and the timelike Liouville contributions. Through the lens of the Liouville-$H_3^+$ WZW correspondence \cite{RibaultTeschner}, these cancellations can be associated to similar phenomena that take place in the Minimal Liouville Gravity and in superstring theory on AdS$_3\times S^3\times M_4$. In fact, this gives us the chance to discuss the relation between our solutions and some other string backgrounds recently discussed in the literature: The first observation is that relation (\ref{relationen}) is exactly the one holding between the KM level and the Liouville parameter in the $H_3^+$ WZW-Liouville correspondence. Next, we observe that when going from the timelike to the spacelike Liouville theory one considers the analytic continuation $b\to {ib=q}$, so that the point $q=\frac{1}{\sqrt 2}$ corresponds to the Liouville parameter $
{b}^{-2}=-2$. This yields the relation ${b}=-\frac{1}{2{b}}$, which means that the Liouville screening operator $e^{\sqrt{2}b\phi (z,\bar z)}$ happens to coincide with the degenerate field $e^{-\frac{1}{\sqrt {2}b}\phi (z,\bar z)}$, with $\phi =iX^0$ being the spacelike Liouville field. This means that, for $\kappa =4$, the two additional degenerate operators standing in the Liouville side of the $H_3^+$ WZW-Liouville correspondence \cite{RibaultTeschner} reduce to dimension-($1,1$) screening operators. This explains why the correlation functions of the AdS$_3\times \text{Liouville}_{c=-2}$ piece manifestly show the same type of cancellation that takes place between the spacelike and the timelike DOZZ structure constants. This coincidence between screening and degenerate fields at $b^2=-\frac 12$ has been early discussed in relation to the $H_3^+$ WZW-Liouville correspondence in \cite{Nakayama}, where this was obtained as the $\kappa \to 0$ limit of the $SL(2.\mathbb{R})_{\kappa }$ WZW model. As observed in \cite{Nakayama}, this can be thought of as the fixed point of different duality symmetries, including the Fateev-Zamolodchikov-Zamolodchikov duality. This partially explains the remarkable simplification we observed at $
\kappa =4 , \, q=\frac{1}{\sqrt 2}$. In addition, there are other backgrounds for which special features are expected to occur. In particular, it would be interesting to explore the case with $\kappa =3 , \, q=1$. In terms of the $H_3^+$ WZW-Liouville correspondence, the latter solution can be related to backgrounds of the form Liouville$_{c=25} \times \mathbb{R}$, with $\mathbb{R}$ being time. While we were finishing our paper, a very interesting work \cite{Balthazar:2023oln} appeared in which the latter backgrounds are studied in detail. Also there the authors study a series of marginal deformations and their interpretation as time-dependent backgrounds of string theory. It would be very interesting to explore the relationship between the results of \cite{Balthazar:2023oln} and our analysis.


\end{document}